\documentclass[a4paper,prl,twocolumn,superscriptaddress,longbibliography]{revtex4-1}

\usepackage{amssymb}
\usepackage{amsmath}
\usepackage{epsfig}
\usepackage{color}
\usepackage{graphics, graphicx}
\usepackage{bbold}
\usepackage{psfrag}
\usepackage{mathcomp}
\usepackage{subfigure}
\usepackage{verbatim}
\usepackage{color}
\usepackage{blindtext}
\usepackage{float}
\usepackage[utf8]{inputenc}
\usepackage[T1]{fontenc}
\usepackage[colorlinks,citecolor=blue]{hyperref}

\usepackage[normalem]{ulem}

\def\GJ{\textcolor{black}}

\begin{document}
%\date{\today}

%\title{Non-Hermitian topology and anomalous skin modes in dissipative quantum systems}
\title{Point-gap topology with complete bulk-boundary correspondence in dissipative quantum systems}

\author{Jian-Song Pan}
\affiliation{Department of Physics, National University of Singapore, Singapore 117542, Singapore}

\author{Linhu Li}
\affiliation{Guangdong Provincial Key Laboratory of Quantum Metrology and Sensing $\&$ School of Physics and Astronomy, Sun Yat-Sen University (Zhuhai Campus), Zhuhai 519082, China}

\author{Jiangbin Gong}
\email{phygj@nus.edu.sg}
\affiliation{Department of Physics, National University of Singapore, Singapore 117542, Singapore}

\begin{abstract}
\GJ{The spectral and dynamical properties of dissipative quantum systems, as modeled by
a damped oscillator in the Fock space, are investigated from a topological point of view.
Unlike a physical lattice system that is naturally under the open boundary condition, the bounded-from-below nature of the Fock space offers a unique setting for understanding and verifying  non-Hermitian skin modes under semi-infinity boundary conditions that are elusive in actual physical lattices.
A topological characterization based on the complex spectra of the Liouvillian superoperator is proposed and the associated complete set of  topologically protected skin modes
can be identified, thus reflecting the complete bulk-boundary correspondence of point-gap topology generally absent in realistic materials.  Moreover, we discover anomalous skin modes with exponential amplification even though the quantum system is purely dissipative.  Our results indicate that current studies of non-Hermitian topological matter can greatly benefit research on quantum open systems and vice versa. }
\end{abstract}

\maketitle
%%%%%%%%%%%%%%
%%%%%%%%%%%%%%
{\color{blue}\emph{Introduction}}.--
The principle of bulk-boundary correspondence (BBC) in topological phases of matter has inspired many innovations in electronics and spintronics~\cite{bernevig2013topological,asboth2016short,ortmann2015topological,shun2018topological}.
For example,  gapless edge modes (in Hermitian systems) under the open boundary condition (OBC) can be predicted by some topological invariants defined for the bulk eigenstates under the periodic boundary condition (PBC)~\cite{hasan2010colloquium,qi2011topological,chiu2016classification,armitage2018weyl,nayak2008non}.
By contrast, in non-Hermitian topological systems, the conventional BBC is  violated in general, where non-reciprocal hopping, asymmetric gain and loss    may pump all eigenmodes to the system's open boundaries, a feature known as the non-Hermitian skin effect (NHSE)~\cite{yao2018edge,kunst2018biorthogonal,xiong2018does,li2019geometric,alvarez2018topological,shen2018topological,yin2018geometrical,yokomizo2019non,song2019non,okuma2019topological,kawabata2019symmetry,bergholtz2019exceptional,lee2019anatomy,song2019non,jin2019bulk,hofmann2020reciprocal,helbig2020generalized,edvardsson2019non,ezawa2019non,luo2019higher,borgnia2020non,xiao2020non,zhang2020bulk,okuma2020topological,zhang2020correspondence,lee2020unraveling,mu2020emergent,lee2020ultrafast,yi2020non}.
As a consequence of NHSE, the bulk states are highly sensitive to boundary conditions, with both their spatial profiles and spectra under the PBC much different from that under the OBC~\cite{krause1994bounds,yao2018edge,gong2018topological,kunst2018biorthogonal,li2019geometric,herviou2019defining,budich2020non,li2020critical,herviou2019defining,xiong2018does,koch2020bulk,lee2016anomalous,deng2019non}.
Recently, different generalized versions of BBC based on redefining the bulk topological indices to incorporate the impact of NHSE have been proposed ~\cite{kunst2018biorthogonal,yao2018non,edvardsson2019non,kunst2019non,yao2018edge,yokomizo2019non,song2019non,herviou2019defining,herviou2019entanglement,
borgnia2020non,zirnstein2019bulk,lee2019anatomy,imura2019generalized}.

As an interesting and important development, it is found that NHSE is not just adding complications to the conventional BBC physics -- NHSE itself is also a topological effect associated with {energetics around point-shape gaps~\cite{point_gap}},  namely, the spectral winding on the complex energy plane with a reference energy~\cite{okuma2020topological,zhang2020correspondence}. The point-gap topology is classified by the homotopy group of general linear group $GL(\mathbb{C},n)$ and is unique for non-Hermitian systems~\cite{gong2018topological,kawabata2019symmetry,bergholtz2019exceptional}.  As such, localized modes due to NHSE can be understood as another type of topological edge modes, and the concept of BBC is now generalized to also address the relationship between NHSE and the spectral winding. However, the complete correspondence between edge modes under OBC and non-trivial point-gap topology under PBC is still absent~\cite{gong2018topological,okuma2020topological}.
%\textcolor{blue}{(Because our logic of this paragraph is that: 1) BBC for non-Hermitian topology defined for Bloch states %are generally broken; 2) Interestingly, NHSE is associated with point-gap topology; 3)However, the complete BBC of point-%gap topology is also absent. Then we should rewrite with "however" and "is absent as well" to make the logic flow %smooth.)}

%Nevertheless, such a correspondence between edge modes and the spectral flow is generally absent under the OBCs
%%\cite{gong2018topological,okuma2020topological}.
%This statement seems incorrect, because NHSE is the result under the OBCs.

\begin{figure}[tbp]
\includegraphics[width=8.5cm]{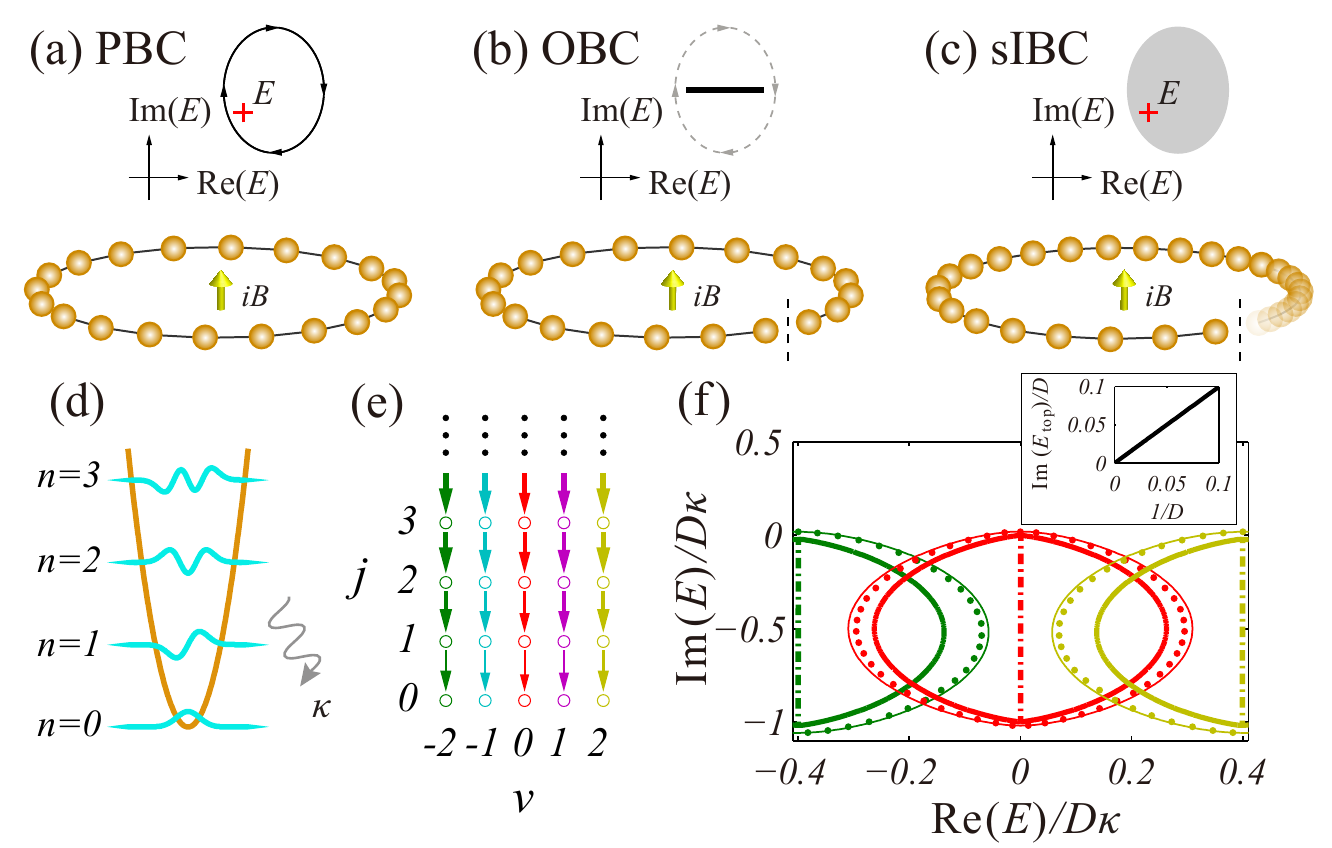}
\caption{(a)-(c) Illustration of configurations and spectra of a system with point-gap topology under different boundary conditions (the blueprint is the HN model). Under OBC the spectra shrink into a line without an area.  Under sIBC the skin modes at each reference energy $E$ inside the PBC spectral loop emerge. (d) A quantum damped oscillator model. (e) Illustration of Fock-space lattice.
(f) Energy spectra under PBC calculated with numerical diagonalization (dots) and predicted with semi-analytical calculation (thin solid curves) for $D=50$. For $D\rightarrow\infty$, the spectral profiles become the bold solid curves. The dash-dotted vertical lines represent the OBC spectra for $D\rightarrow\infty$ (homogeneously distributed on these lines when $D$ is finite). The colors specify the spectra of different 1D chains, as shown in (e). Insert: the finite-size scaling of the topmost point of PBC spectra loop for chain $\nu=0$ (red), indicating that all cases approach the same point $E=i\kappa$. The parameters are set at $\hbar\omega=1$ and $\kappa=0.1$ with dimensionless units. }
\label{fig:spectra}
\end{figure}

Of particular interest is hence the semi-infinity boundary condition (sIBC), under which a system is only open in one direction, but infinitely extended in the other direction.  Indeed, it is the sIBC that makes it possible to have a complete story of BBC associated with point-gap topology, e.g., what is the implication of different spectral winding behaviors for skin modes.  To appreciate why sIBC is important,  it is illustrative to first consider a PBC point-gap topology, e.g., that of the Hatano-Nelson (HN) model~\cite{hatano1997vortex,hatano1998non}, described by a simple Hamiltonian $H_{\text{HN}}= \sum_{j} [ t_{\rm {R}}\hat{c}_{j+1}^{\dagger}\hat{c}_{j}+t_{\rm{L}}\hat{c}^{\dagger}_{j}\hat{c}_{j+1}]\ (t_R\ne t_L)$.  Without breaking any symmetry, the {PBC} spectral loop
is turned into a straight line on the real axis under OBC [see Fig.~\ref{fig:spectra}(a) and (b)].
%which breaks down the doctrine that the changes of boundaries doesn't have significant effect on the bulk states, and then %the true sense of the conventional BBC is absent [see Fig.~\ref{fig:spectra}(a) and %%(b)]~\cite{gong2018topological,okuma2020topological,zhang2020correspondence,borgnia2020non,bergholtz2019exceptional}.
One then inquires about the spectral topology for sIBC [see Fig. \ref{fig:spectra}(c)].  Interestingly, the spectral loop destroyed by OBC can be retrieved by sIBC, with the following  qualitative reason:  Under PBC the asymmetric hopping can be understood as an imaginary gauge potential~\cite{imaginary_gauge_field}; with OBC capable of gauging away this imaginary potential but sIBC cannot.
Thus,  sIBC is peculiar because, on the one hand the system is of infinite size with spectral winding on the complex plane and on the other hand there is an edge to manifest the spectral winding. Indeed, both the existence and degeneracy of such edge modes at a given reference energy inside the spectral loop can now be
predicted by a spectral winding number~\cite{gong2018topological,okuma2020topological}.
 It should be stressed that such a sIBC scenario emerges only in the true infinite-size limit.  In mathematics, this type of modes was predicted in the framework of infinite Toeplitz matrix decades ago~\cite{reichel1992eigenvalues, bottcher2005spectral}. Nevertheless, the sIBC configuration cannot be realized in any realistic lattice system since one literally needs a system of infinite size to start with.

In this work,  we find that the BBC under sIBC offers important insights when investigating dissipative quantum systems.  Specifically,  upon mapping Fock states to single-particle states on a lattice, the mapped system is guaranteed to be under sIBC and the spectra of the Liouvillian superoperator can form spectral winding with respect to a twisting angle parameter.
Similar to the Hatano-Nelson model, the continuum of sIBC spectra enclosing the OBC spectra inside the spectral loops is identified.  This not only provides abundant opportunities towards understanding the system dynamics but also represents the first faithful realization of the complete BBC of point-gap topology.
For a damped harmonic oscillator as an example, we find anomalous skin modes, which show exponential amplification rather than attenuation, even though the system under consideration is purely dissipative.  Our work thus advocates a feasible platform to experimentally investigate the BBC under sIBC and reveals previously unknown possibilities in quantum dissipative systems.  Indeed, even coherent states of light can be regarded as point-gap topology protected skin modes in the Fock space. This work differs from other studies of quantum open systems in connection with non-Hermitian topological physics~\cite{song2019nonlindblad, lieu2020tenfold, haga2020liouvillian, gneiting2020unraveling, longhi2020unraveling} in that we exploit sIBC and examine edge modes in the Fock space.

{\color{blue}\emph{The model}}.--Consider a quantum harmonic oscillator damped through Markovian processes [see Fig.~\ref{fig:spectra}(d)], whose dynamics is governed by the following Lindblad master equation,
\begin{equation}\label{eq:Lindblad}
\frac{\partial\hat{\rho}}{\partial t}=i[\hat{\rho},\hat{H}]+\kappa(2\hat{a}\hat{\rho}\hat{a}^{\dagger}-\hat{a}^{\dagger}\hat{a}\hat{\rho}-\hat{\rho}\hat{a}^{\dagger}\hat{a})/2=\mathcal{L}\hat{\rho},
\end{equation}
where $\hat{\rho}$ is the density operator, $\hat{a}$ ($\hat{a}^\dagger$) the annihilation (creation) operator (e.g., regarding photons in a quantum optics setting), $\kappa$ the damping coefficient, and $\hat{H}=\hbar\omega(\hat{a}^{\dagger}\hat{a}+1/2)$ the system Hamiltonian. In above we have also defined $\mathcal{L}$ as the Liouvillian superoperator. Note that this is a textbook model~\cite{rivas2012open,ficek2014quantum,honda2010spectral,fujii2013quantum} but here we shall treat it from a different angle based on the generalized BBC, with unexpected results.

Under the Fock state representation, namely, $|m\rangle\langle n|$ with integers $m,n\geq 0$, the above operators can be expressed as $\hat{\rho}=\sum_{m,n=0}^{\infty}\rho_{mn}|m\rangle\langle n|$, $\hat{a}=\sum_{m,n=0}^{\infty}a_{mn}|m\rangle\langle n|$ with $a_{mn}=\sqrt{m+1}\delta_{m+1,n}$, and $\hat{H}=\sum_{m,n=0}^{\infty}H_{mn}|m\rangle\langle n|$ with $H_{mn}=(m+1/2)\delta_{m,n}$. The Lindblad equation can be also rewritten as a Schrodinger-like equation $i\partial_{t} |\rho\rangle=\mathcal{H}|\rho\rangle$ with $|\rho\rangle=\sum_{m,n=0}^{\infty}\rho_{mn}|mn\rangle$ in the two-mode space $\{|mn\rangle, m, n\geq 0\}$~\cite{navarrete2015open}, and
\begin{equation}\label{eq:NHHam}
\mathcal{H}=H\otimes I-I\otimes H^{T}+i\kappa(2\hat{a}\otimes\hat{a}^{\ast}-\hat{a}^{\dagger}\hat{a}\otimes I-I\otimes\hat{a}^{\dagger}\hat{a})/2.
\end{equation}
Time evolution of $\hat{\rho}$ is then given by $|\rho(t)\rangle=e^{-i\mathcal{H}t}|\rho(0)\rangle$.
%Then the eigenstates and eigenvalues of $\mathcal{H}$, which are focused on in this paper, encode the dynamics once the initial state is specified.

Because the Schrodinger-like equation is understood in the two-mode basis, mapping the Fock states to single-particle states on a lattice~\cite{wang2016mesoscopic,cai2019topological,lee2020many} yields a two-dimensional (2D) lattice labeled by $(m,n)$. As the only hopping term $2i\kappa \hat{a}\otimes \hat{a}^{\ast}$ couples site $(m,n)$ to sites $(m\pm1,n\pm 1)$, the 2D lattice is actually formed by a series of decoupled one-dimensional (1D) chains under sIBC. The decoupled chains can be specified with photon number differences $\nu=(m-n)$ [see Fig.~\ref{fig:spectra}(e)], i.e. $\mathcal{H}=\bigoplus_{\nu=-\infty}^{\infty}\mathcal{H}_{\nu}$ with
\begin{equation}\label{Eq:1D_chain}
\mathcal{H}_{\nu}=\sum_{j=0}^{\infty}(h_{\nu j}\hat{c}_{\nu j}^{\dagger}\hat{c}_{\nu j}+t_{\nu j}\hat{c}_{\nu j}^{\dagger}\hat{c}_{\nu j+1}),
\end{equation}
where $\hat{c}_{\nu j}$ ($\hat{c}_{\nu j}^{\dagger}$) are the annihilation (creation) operators at site $(j+\frac{|\nu|+\nu}{2},j+\frac{|\nu|-\nu}{2})$, $h_{\nu j}=\nu\omega-i\kappa(2j+|\nu|)/2$, and $t_{\nu j}=i\kappa\sqrt{(j+1)(j+|\nu|+1)}$. $\mathcal{H}_{\nu}$ characterizes a 1D non-Hermitian chain with unidirectional hopping like the HN model under sIBC~\cite{hatano1996localization, hatano1997vortex, hatano1998non}, but no longer possessing the translation symmetry. Hence this model goes beyond Toeplitz-matrix descriptions~\cite{translation,reichel1992eigenvalues, bottcher2005spectral} and it remains to investigate if $\mathcal{H}$ displays spectral winding under  PBC and consequently edge spectra under its natural sIBC.

{\color{blue}\emph{Spectral properties under fictitious PBC}}.-- Let us first consider a fictitious PBC situation by truncating the dimension of Fock space at $(D+1)$ and linking site $D$ to site $0$. This leads to
\begin{equation}
\mathcal{H}_{\text{PBC},\nu}^{(D)}=\sum_{j=0}^{D}(h_{\nu j}\hat{c}_{\nu j}^{\dagger}\hat{c}_{\nu j}+t_{\nu j}\hat{c}_{\nu j}^{\dagger}\hat{c}_{\nu\text{mod}(j+1,D)}),
\end{equation}
and $\mathcal{H}_{\text{PBC},\nu}=\lim_{D\rightarrow\infty}\mathcal{H}_{\text{PBC},\nu}^{(D)}$.  % The construction of such PBC has no experimental %significance since Fock space is infinitely extended.
%Practically, the Fock-space lattice is infinitely extended and the eigenfunctions is restricted by the sIBC, which only requires the eigenfunctions to terminate (to %terminate or asymptotically tend to zero) in one (the other) direction. In contrast, the conventional OBC of realistic materials requires the eigenfunctions to %terminate in both directions. Nevertheless, the information of spectra under these three types of boundary condition provide different perspectives to our model, %and are all discussed below.
%{\color{blue}\emph{\text{PBC} spectra and winding number}}.--
 The PBC spectra in the infinite-size limit $D\rightarrow\infty$ can then be found from how the spectra scale with dimension $D$.

The characteristic equation of $\mathcal{H}_{\text{PBC},\nu}^{(D)}$ is equivalent to
\begin{equation}\label{eq:characteristic}
R_{\nu}^{(D)}(E)=\Pi_{j=0}^{D}\frac{E-h_{\nu j}}{t_{\nu j}}=\Pi_{j=0}^{D}\gamma_{\nu j}(E)=1,
\end{equation}
i.e. $\ln(R_{\nu}^{(D)}(E))=0$. Employing the approximation $\sum_{j=0}^{D}\ln(|\gamma_{\nu j}(E)|)\approx\int_{0}^{D}dx\ln(|\gamma_{\nu x}(E)|)$ when $D\gg1$, one obtains
\begin{equation}\label{eq:scaled_R}
\begin{split}
D^{-1}\ln |R_{\nu}^{(D)}(E)|\approx&f(1+\tilde{E}/D,1+|\nu|/D,1+1/D)  \\
&-f(\tilde{E}/D,|\nu|/D,1/D),
\end{split}
\end{equation}
where $\tilde{E}=[E+(i|\nu|\kappa-\nu\omega)]/i\kappa$ and $f(A,a,b)=A_{x}\ln|A|-A_{y}\arg(A)-(a\ln a+b\ln b)/2$ with complex variable $A=A_x+iA_y=\tilde{E}/D, 1+\tilde{E}/D$ and real variables $a=|\nu|/D, 1+|\nu|/D$ and $b=1/D, 1+1/D$. %Thus, the profile of PBC spectra is determined by the equation $\ln(|R_{\nu}^{(D)}(E)|)=0$.
%Under the large-$D$ approximation here, the eigenvalue equation $\ln(|R_{\nu}^{(D)}%(E)|)=0$ indicates that there is a spectral loop structure (see Supplementary Material~\cite{suplementary} for details), in excellent agreement with
Our results from direct numerical diagonalization and Eq.~(\ref{eq:scaled_R}) agree well and both indicate that when $D\gg 1$, the spectra are loops on the complex plane [see Fig.~\ref{fig:spectra}(f)].  Further, our theoretical expressions here
also indicate that in the limit of $D\rightarrow\infty$, $a\ln(a)+b\ln(b)$ term vanishes and then the eigenvalue equation depends only on the scaled energies $E/D$ and chain indices $\nu/D$.  Thus, the radius of any spectral loop diverges linearly with $D$. The requirement that $\arg(R_{\nu}^{(D)}(E_{\text{PBC},\nu l}^{(D)}))=2l\pi$ determines the actual distribution of PBC eigenvalues $\{E_{\text{PBC},\nu l}^{(D)}|0\leq l\leq D\}$ on a spectral loop.  Interestingly, the eigenvalues for any $\nu$th chain tend to uniformly distribute on its spectral loop~\cite{PBCspectra}.

From the classification theory of non-Hermitian topology, a 1D system is topologically nontrivial provided that
its complex spectra enclose certain point gaps giving nonzero winding numbers~\cite{gong2018topological,okuma2020topological,zhang2020correspondence,bergholtz2019exceptional}.
For our 1D system here that does not have the translation invariance, one may further characterize the spectral winding
by introducing the twisted boundary condition (TBC) $t_{\nu D}\rightarrow e^{i\theta}t_{\nu D}$.  For $\theta=0\rightarrow 2\pi$, from Eq.~(\ref{eq:characteristic}) and that
$\arg(R_{\nu}^{(D)}(E_{\text{PBC},\nu l}^{(D)}))=2l\pi$ determines PBC eigenvalues,
one expects the spectrum to have finished one complete rotation. We then proceed to define spectral winding numbers as
\begin{equation}
W_{\nu}(\Omega)=\lim_{D\rightarrow \infty}\frac{1}{2\pi i}\int_{0}^{2\pi}d\theta\frac{\partial}{\partial\theta}\text{Tr}[\ln(\mathcal{H}_{\nu}^{(D)}(\theta)-\Omega)],
\end{equation}
where $\mathcal{H}_{\nu}^{(D)}(\theta)$ is the TBC Hamiltonian and $\Omega$ is the reference energy on the complex plane~\cite{gong2018topological}. If $\Omega$ is inside (outside) the spectral loop, then $W_{\nu}(\Omega)$ equals $-1$ ($0$).  In the following we focus on the implications of nontrivial spectral windings.
% a continuum of skin modes fully filling up the spectral loop of $\mathcal{H}%_{\nu}^{(D)}(\theta)$ is expected to exist under appropriate boundary condition.

\begin{figure*}[tbp]
\includegraphics[width=17.5cm]{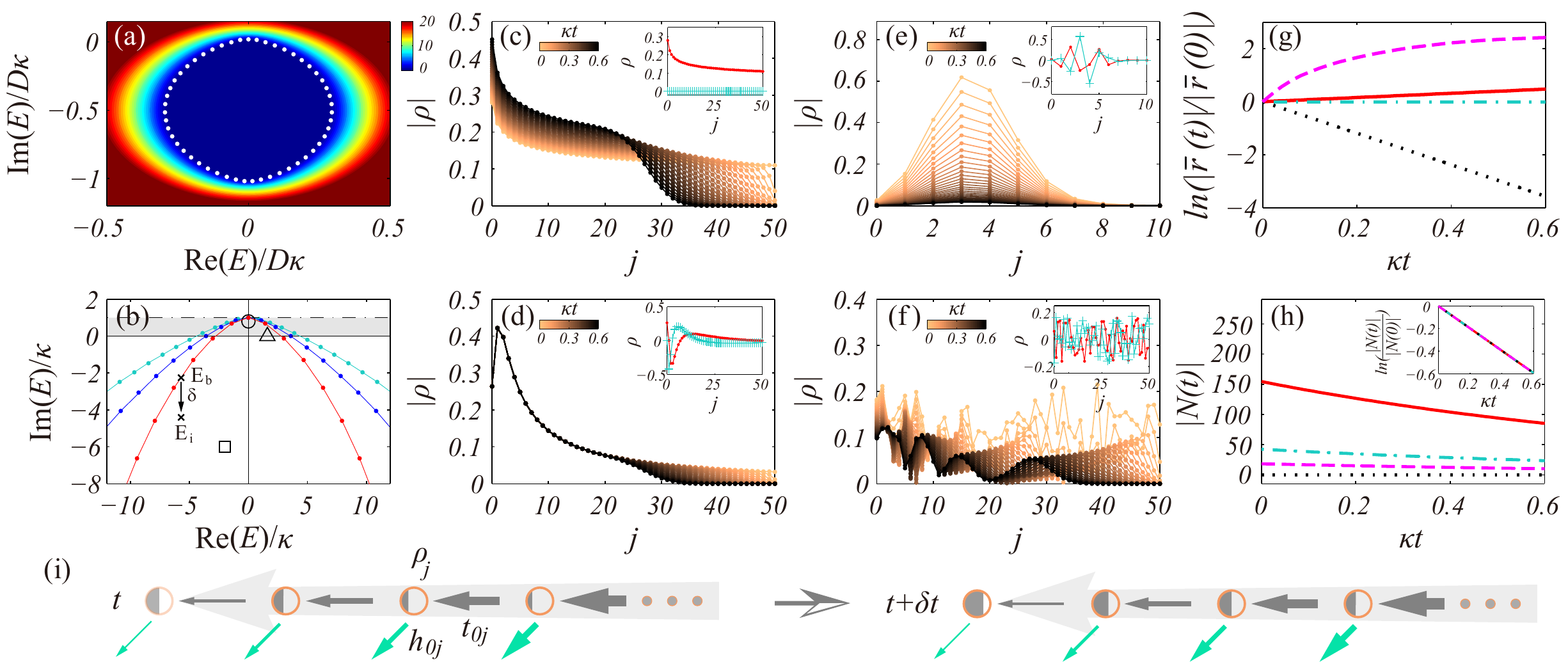}
\caption{(a) Function $|R|$ shown via $\ln (|R_{0}^{(50)}(E)|+1)$ (the plotted function is chosen to view the border clearly with different colors), with the white dots depicting the PBC spectral loops with
$R_{0}^{(50)}(E)=1$.  (b) The upper ridges of spectral loop with different $D$, for $\nu=0$ and $D=50$ (red dots), $200$ (blue dots) and $400$ (cyan dots), all crossing the same point $E=i\kappa$, and finally becoming the horizontal line $E=i\kappa$ for $D\rightarrow\infty$ (dash-dotted line). The shadow regime represents the continuum of anomalous skin modes with positive imaginary parts.
%It is seen that any energy value (e.g. $E_{i}$) inside the loops become sIBC eigenvalues when $D\rightarrow\infty$~\cite{suplementary}.
(c)-(f) Evolution of typical skin modes above (circle), on (triangle) and below (square) the real axis, and that for a random initial state, with a truncation at $j=50$ and initial states in the inserts (red dots: real part; cyan crosses: imaginary part). (g) The average amplitude of the first five sites, $\bar{\rho}(t)=5^{-1}\sum_{j=0}^{4}\rho_{j}(t)$, where the red solid, cyan dash-dotted, black dotted, magenta dashed curves correspond to the edge modes in (c)-(f), respectively. (h) Evolution of particle number $N(t)=\text{Tr}(\rho(t)\hat{a}^{\dagger}\hat{a})$ (%although $\rho$ is treated as a density matrix, it may not satisfy the requirements as a density matrix, such that even $N(t)$ may be a complex number, as discussed in the main text)
with the same line convention as in (g). The insert shows the logarithm plot of $|N(t)/N(0)|$, indicating that the total particle number does show exponential damping in all the cases.
(i) Schematic of the physics behind the amplification dynamics of anomalous skin modes.
Other system parameters used in figure panels (a)-(h) are set to be the same as in Fig.~1.
 }
%(i) The schematic of the amplification dynamics of anomalous edge modes. The other parameters are set to be the same with Fig.~\ref{fig:spectra}.}
\label{fig:irregularSP}
\end{figure*}

{\color{blue}\emph{OBC spectra}}.--For the sake of comparison with the spectra under sIBC, we next investigate the spectra under OBC by artificially truncating the Fock space at the occupation number $D$. The OBC spectra and corresponding eigenstates are given by $\mathcal{E}_{\text{OBC}}=\{E_{\text{OBC},\nu l}=\nu\omega-i(2l+|\nu|)\kappa/2|$ and $|\rho_{\text{OBC},\nu l}\rangle\propto\sum_{j=0}^{l}(-1)^{j}\sqrt{C_{l+|\nu|+j}^{j}C_{l+1}^{j}}|\nu(l-j)\rangle$, with $C_l^j=l!/j!(l-j)!$, respectively. For each individual decoupled 1D chain, the obtained eigenvalues scaled by dimension $D$ are found to distribute homogeneously on straight lines inside its PBC spectral loop, {which indicates that the point-gap topology has been destroyed} [see Fig.~\ref{fig:spectra}(f)]. In particular, the imaginary parts of the spectra $\text{Im}(E_{\text{OBC},\nu l})\leq 0$, indicating that the dynamical evolution of these states
shows attenuation, whereas only the state $E_{\text{OBC},00}=0$ has no attenuation since the Fock space is bounded from below.  %The OBC modes and their dual left eigenmodes constitute a bi-orthogonal complete basis set.%
 {Although the eigenstates are not completely localized at around the boundary due to the lack of translation symmetry in the Fock state representation, the results here are consistent with the known aspects of NHSE as a consequence of point-gap topology (see Supplementary Material \cite{suplementary} for details)~\cite{gong2018topological,okuma2020topological,zhang2020correspondence}. Results here also agree with previous studies on the spectra of a damped  oscillator~\cite{englert2002five,rivas2012open,ficek2014quantum,honda2010spectral,fujii2013quantum,albert2014symmetries}.}

{\color{blue}\emph{sIBC spectra and edge modes}}.-- Let us now investigate the more interesting spectra from the eigenequation $\mathcal{H}|\rho\rangle=E|\rho\rangle$ under sIBC.  The sIBC only requires $\lim_{j\rightarrow\infty}\rho_{\nu j}=0$ (instead of $\rho_{\nu j}=0$ for specific $j$ under the OBC), where $\rho_{\nu j}=\langle \nu j|\rho\rangle$.
% Then, the sIBC spectra $\mathcal{E}_{\text{sIBC}}\supset \mathcal{E}_{\text{OBC}}$ by construction.
We will show below that $\mathcal{E}_{\text{sIBC}}$ is a continuum of edge spectra inside the PBC spectral loop, thus also including the OBC spectra.

Under sIBC one expects to see solutions non-vanishing in any finite regime and converging to zero only in the infinite-size limit. Interestingly, these solutions satisfy $\rho_{\nu j+1}=R_{\nu}^{(j)}(E)\rho_{\nu 0}$ from the eigenequation and then the boundary condition is equivalent to $\lim_{j\rightarrow\infty}|R_{\nu}^{(j)}(E)|=0$.  Note that the condition $|R_{\nu}^{(D)}(E)|=1$ for the $\nu$th decoupled 1D chain yields its PBC spectral loop and the solutions of $|R_{\nu}^{(D)}(E)|<1$ ($|R_{\nu}^{(D)}(E)|>1$) are located inside (outside) this loop. Thus, the condition $\lim_{j\rightarrow\infty}|R_{\nu}^{(j)}(E)|=0$ requires the sIBC solutions must distribute inside the PBC spectral loop under $D\rightarrow\infty$. From numerics, given the scaled function $D^{-1}\ln |R_{\nu}^{(D)}(E)|=-c\in (-\infty,0)$ inside the spectral loop from Eq.~(\ref{eq:scaled_R}),  we always have $\lim_{j\rightarrow \infty} |R_{\nu}^{(j)}(E)|=\lim_{j\rightarrow \infty}\exp(-jc)=0$ for any fixed scaled energy $\tilde{E}/D$.  This observation implies the sIBC spectra tend to fully fill up the PBC spectral loop (in the sense of the scaled spectrum) [see Fig.~\ref{fig:irregularSP}(a)].  Certainly, the language of scaled spectra blurs fine details of the boundary of sIBC spectra, because eigenvalues with small magnitudes like $E=i\kappa$ and $2i\kappa$ are all scaled down to $0$ in the infinite-size limit. For this reason we still need to further check the precise boundary of sIBC spectra without scaling. {As shown in Supplementary Material~\cite{suplementary}, for $D\rightarrow\infty$, any energy value inside the spectral loop are indeed valid eigenvalues under the sIBC.  It can be hence concluded that the spectra of the Liouvillian superoperator  $\mathcal{L}$ naturally under sIBC  are the continuum bounded by the PBC spectral loops, a result unnoticed in previous studies~\cite{englert2002five,rivas2012open,ficek2014quantum,honda2010spectral,fujii2013quantum,albert2014symmetries}.}

{\color{blue}\emph{Anomalous skin modes}}.-- Here we focus on the $\nu=0$ chain, as it corresponds to the diagonal components of the density matrix. Considering $|R_{0}^{(D)}(i\kappa)|=1$ as the eigenvalue equation under PBC, the PBC spectral loop is found to cross the  same point $E=i\kappa$ for sufficiently large $D$ (see  inset of Fig.~\ref{fig:spectra}h). On the other hand, since the radius of PBC spectral loop is proportional to $D$, the upper ridge of the PBC spectral loop becomes a horizontal line $\text{Im}(E_{b}^{(0)})=\kappa>0$ on the complex plane when $D\rightarrow \infty$. This is shown in Fig.~\ref{fig:irregularSP}(b).  Remarkably, unlike the OBC case, the sIBC spectra, which must fill up the PBC spectra loop as shown above, may go above the real axis and hence predict anomalous skin modes with positive imaginary parts.  That is, the sIBC spectra are not only qualitatively different from the OBC spectra,  but also predict, counter-intuitively, amplification dynamics even though the damped quantum system is purely dissipative.

To confirm the existence of such anomalous skin modes, Fig.~\ref{fig:irregularSP}(c)-(f) presents computational results for verification.
A finite-size truncation has to be introduced for the used initial states.  Therefore the reliable time duration for such numerical experiments does depend on how fast the skin modes can be amplified.  That is, exponential amplification of anomalous skin modes is indeed observed
within a certain period to exclude the inevitable finite-size effects from numerical truncation [see (c) and red solid line in (g)]. By contrast, the dynamics of normal skin modes with zero and negative imaginary parts shows steady time evolution and exponential attenuation, respectively [see (d) and (e), and cyan dash-dotted and dotted curves in (g)].  As a comparison, the evolution of a rather random localized mode with irregular variation is also shown [see (f) and magenta dashed curve in (g)].

Further discussions are needed to digest the amplification dynamics of the anomalous skin modes discovered here.  As shown in the insert of Fig.~\ref{fig:irregularSP}(h), the total particle number is always damped as $e^{-\kappa t}$.  The photon number increase in the amplification of the anomalous edge mode must come from sites with large-photon-number.  This is confirmed by the large-$j$ part of Fig.~\ref{fig:irregularSP}(c).
The following picture hence emerges. The probability density flows from large-photon-number (upper) sites to small-photon-number sites (lower sites), and pile up at the latter,  with the total photon number still damped.
More specifically, with the initial amplitude $\rho_{j}=\langle 0j|\rho\rangle$, the dynamical evolution over a short time $\delta t$ leads to $\rho_{j}(t+\delta t)=\langle 0j|e^{-i\mathcal{H}\delta t}|\rho\rangle\approx [1-j\kappa\delta t+(j+1)\kappa\delta t \rho_{j+1}/\rho_{j}]\rho_{j}$. This
 yields  $\rho_{j}(t+\delta t)>\rho_{j}$ once $\rho_{j+1}/\rho_{j}>j/(j+1)$. For the anomalous skin modes, $\rho_{j+1}/\rho_{j}=(j+\tilde{E})/(j+1)$ with $\tilde{E}\in (0,1)$. Hence its amplitude is amplified. As shown in Fig.~\ref{fig:irregularSP}(i), for each site, the loss of probability by the diagonal term $h_{0j}$ is over offset by the input by the off-diagonal term $t_{0j}$, leading to a net input flow from upper sites.
%This phenomenon is completely counterintuitive since the model we consider is purely dissipative.

As a final remark, we note that the eigenvectors of  the Liouvillian superoperator defined above are not necessarily well-defined density matrices because their entries might be negative. Furthermore,  the trace of these eigenvectors may not be normalizable because they form a continuum under sIBC.   Fortunately, the anomalous skin modes of the $\nu=0$ chain on the imaginary axis (e.g. the mode denoted with circle in Fig.~\ref{fig:irregularSP}(b)) always have real positive entries  [see the insert of Fig.~\ref{fig:irregularSP}(c)], thus indeed experimentally relevant to situations where the preparation of an initial density matrix is truncated at a sufficiently large system size, as confirmed by our numerical experiments discussed above.   %While the complete preparation of the system into the anomalous edge modes may be hard, it is possible to observe this counterintuitive %phenomenon up to an arbitrarily large cutoff in experiment, which may inspire new applications in quantum techniques.

{\color{blue}\emph{Conclusion}}.--Non-Hermitian point-gap topology is shown to be an important aspect of the spectra of the Liouville superoperator associated with dissipative quantum systems such as a damped oscillator. Owing to the semi-infinite boundary condition of the Fock space, the complete set of skin modes protected by the spectral winding topology are found, upon mapping the Fock space onto a lattice.   The results go beyond a mapping because the mapped system does not have translational invariance and yet a generalized bulk-boundary correspondence still applies.    The theoretical framework established in this Letter not only offers a unique angle for understanding quantum optics problems, but also uncovers  a new playground to look into non-Hermitian systems.    One key finding here is the anomalous skin modes with positive imaginary parts and hence amplification in time, in a purely dissipative system.  {Furthermore, if one revisits the annihilation operator with the current framework, even the celebrated coherent states of light are found to be nothing but the continuum of skin modes protected by some spectral winding (see Supplementary Material~\cite{suplementary} for details). It would be promising to extend this study to include periodic driving, multi-particle damping channels~\cite{ficek2014quantum} or other degrees of freedom (e.g. atoms in damped cavity) to further explore topological physics in non-Hermitian systems.

% Therefore, our work opens a new direction for simulating and exploring topological physics with many-body degrees of freedom.

{\color{blue}\emph{Acknowledgements}}.--The authors wish to thank Dr. Lee Ching Hua for very helpful discussions. J.G.  acknowledges funding support from Singapore National Research
Foundation Grant No. NRF- NRFI2017-04 (WBS
No. R-144-000-378-281). J.-S. P. acknowledges the support from the National Natural Science Foundation of China (Grant No. 11904228) before he joined NUS.

\appendix

\begin{widetext}
\section*{Supplementary material}

In this supplementary material, we present  some details regarding signatures of non-Hermitian skin effect under the open boundary condition (OBC) despite broken translational invariance,  the boundaries of spectral loops under the semi-infinity boundary condition (sIBC), as well as the point-gap topology associated with the annihilation operator.

\section{Signature of Non-Hermitian skin effect}
Nontrivial bulk topology under the PBC usually causes the accumulation of OBC eigenstates at around the open boundaries, known as the non-Hermitian skin effect~\cite{yao2018edge, okuma2020topological, zhang2020correspondence}. For our system treated in the Fock space, there is no translational invariance and the corresponding OBC eigenstate of $E_{D,\nu l}$ is given by $|\rho_{OBC,\nu l}\rangle\propto\sum_{j=0}^{l}(-1)^{j}\sqrt{C_{l+|\nu|+1}^{j}C_{l+1}^{j}}|\nu(l-j)\rangle$ with $C_l^j=k!/j!(l-j)!$. Unlike a typical skin state being localized at around the boundary and decays exponentially in the bulk, the solutions $|\rho_{OBC,\nu l}\rangle$here  tend to localized quite far away from the edge since the function $C_l^j$ peaked at around $j=[l/2]$. For example, for the state $|\rho_{OBC,0D}\rangle$, its density distribution has a peak at around the central point [see Fig.~\ref{fig:skineffect}(a)]. This kind of skin mode behavior is hence a unique consequence of broken translation invariance.  Furthermore, as shown in Fig.~\ref{fig:skineffect}(b), the total density of all OBC eigenstates still shows obvious preference of the boundary, as a clear signature of  non-Hermitian skin effect.  We stress that, although the numerical calculations are done for a finite vaue of $D$, they share the same scaled features.

\begin{figure}[tbp]
\includegraphics[width=12cm]{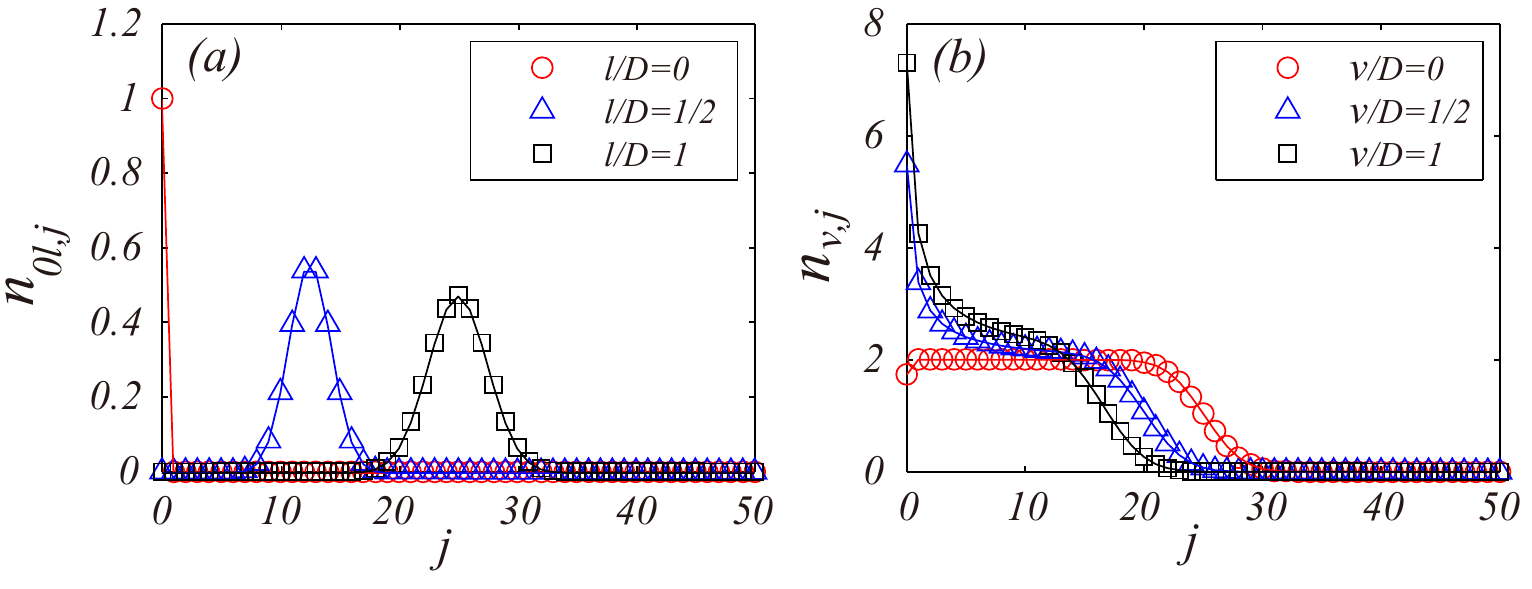}
\caption{Probability density of OBC eigenstates. Typical probability density of one specific OBC eigenstate $n_{\nu l,j}=|\langle \nu j|\rho_{OBC,\nu l}\rangle|^{2}$ is shown in (a). Typical total probability density $n_{\nu,j}=\sum_{l=0}^{D}n_{\nu l,j}$ is shown in (2). The accumulation of total probability density near the lower boundary is a clear signature of non-Hermitian skin effect. The parameters are the same with Fig.~1 in the main text.}
\label{fig:skineffect}
\end{figure}

\section{Boundaries of sIBC spectra}
In the main text it is argued that the sIBC spectra tend to fully fill the PBC spectral loops after scaling with dimension $D$. However, we still need to look into the precise boundaries of sIBC spectra without the scaling. By investigating the behavior of function $|R_{\nu}^{(D)}(E)|$ across a spectral loop, we can conclude that any energies inside the PBC spectral loops actually become the eigenvalues under the sIBC.

Consider then a value of $E_{b}$ precisely on the upper ridge of one spectral loop.  Then $E_{i}=E_{b}-i\delta$ would be located inside the loop [see Fig.~2(b) in the main text].  Then $|R_{\nu}^{(D)}(E_{b})|=1$ and $|R_{\nu}^{(D)}(E_{i})|<1$ according our analysis in the main text.   Their difference would be a finite value and can be expressed by the following:
\begin{equation}\label{eq:RF}
\begin{split}
&|R_{\nu}^{(D)}(E_{b})|-|R_{\nu}^{(D)}(E_{i})|=\int_{0}^{\delta}d\epsilon\frac{\partial|R_{\nu}^{(D)}(E_{i}+i\epsilon)|}{\partial \epsilon}\\
&=\int_{0}^{\delta}d\epsilon|R_{\nu}^{(D)}(E_{i}+i\epsilon)|\Omega_{\nu}^{(D)}(E_{i}+i\epsilon),
\end{split}
\tag{S1}
\end{equation}
where (by simply taking the derivative of the $R$ function defined in the main text)
\begin{equation}
\Omega_{\nu}^{(D)}(E)\equiv\sum_{j=0}^{D}\frac{j+|\nu|/2+\tilde{E}_{y}}{(j+|\nu|/2+\tilde{E}_{y})^{2}+(\tilde{E}_{x}-\nu\omega/\kappa)^{2}},
\tag{S2}
\end{equation}
with $\tilde{E}=E/\kappa=\tilde{E}_{x}+i\tilde{E}_{y}$~\cite{note_integral}. Since there must be large enough $j_0$ satisfying $|j_0+|\nu|/2+\tilde{E}_y|>|\tilde{E}_x-\nu\omega/\kappa|$ and $j_0+|\nu|/2+\tilde{E}_y>0$, we have $\lim_{D\rightarrow\infty}\Omega_{\nu}^{(D)}(E)>\sum_{j=0}^{j_{0}-1}\frac{j+|\nu|/2+\tilde{E}_{y}}{(j+|\nu|/2+\tilde{E}_{y})^{2}+(\tilde{E}_{x}-\nu\omega/\kappa)^{2}}+2^{-1}\sum_{j=j_{0}}^{\infty}\frac{1}{j+|\nu|/2+\tilde{E}_{y}}$. The second term, $\sum_{j=j_{0}}^{\infty}\frac{1}{j+|\nu|/2+\tilde{E}_{y}}$ is divergent, then
 $\Omega_{\nu}^{(D)}(E)$ is also divergent.
On the other hand, considering $\lim_{D\rightarrow \infty}D^{-1}\ln(|R_{\nu}^{(D)}(E)|)<0$ when $E$ is inside the spectral loops from the numerical calculation of Eq. (6) in the main text (see Fig. 2(a) in the main text for the case of $D=50$),  the left hand side of Eq.~(\ref{eq:RF}) must be finite but on the right hand side,  one sees an integral involving divergent $\Omega_{\nu}^{(D)}(E)$.  This observation leads to the conclusion  that $\lim_{D\rightarrow \infty}|R_{\nu}^{(D)}(E_{i}+i\epsilon)|=0$, which is precisely the claim that $E_{i}$ must be an eigenvalue. Therefore, with similar considerations applied to all energy values at the boundary of the PBC spectral loop,  we arrive at the conclusion that the sIBC spectra indeed fully fill up the PBC spectral loop.   Therefore, the sIBC spectra are nothing but the edge mode continuum with nonzero winding numbers referring to their energies inside the PBC spectral loops. This is also vividly reflected in the following example with decreasing $\kappa$.  As $\kappa$ decreases,  the upper ridge of the spectral loop of chain $\nu=0$ (i.e. $E_{y}=i\kappa$ when $D\rightarrow \infty$, see the main text) moves downward.  This suggess that means some reference energy values, initially topologically nontrivial (being inside the loop) regarding the point-gap topology, become trivial. On the other hand, the edge modes corresponding these reference energies disappear, thus yielding topological phase transitions (from nontrivial to trivial, i.e. with to without edge modes) associated with the point-gap topology.  The BBC associated with point-gap topology here is thus manifested as this edge mode continuum.

\section{Point-gap topology of annihilation operator}
In this section, we provide more details about the non-Hermitian point-gap topology of the annihilation operator. Using the same treatment advocated in the main text,   we are able to find that  the celebrated coherent states can be understood as the edge modes protected by som point-gap topology.

Consider then the Fock basis $\{|n\rangle, n\in \mathbb{N}_{0}\}$. The annihilation operator can be written as  $\hat{a}=\sum_{n=0}^{\infty}\sqrt{n+1}|n\rangle\langle n+1|$. By mapping the Fock states onto single-particle states on a  lattice ~\cite{wang2016mesoscopic,cai2019topological,lee2020many}, $\hat{a}$ becomes the following tight-binding model Hamiltonian $\hat{a}=\sum_{j=0}^{\infty}t_{j}\hat{c}_{j}^{\dagger}\hat{c}_{j+1}$, where $\hat{c}_{j}$ ($\hat{c}_{j}^{\dagger}$) is the annihilation (creation) operator at the $j$-th site corresponding to the Fock state $|n=j\rangle$, and $t_{j}=\sqrt{j+1}$ is the hopping strength.
This tight-binding model with spatially inhomogeneous couplings can be seen as  a generalized version of the Hatano-Nelson model~\cite{hatano1996localization, hatano1997vortex, hatano1998non} that possesses nontrivial exceptional topology~\cite{bergholtz2019exceptional}.
Through this mapping, we also obtain a sIBC lattice model as the particle number $n$ naturally ranges from $0$ to infinity.
It is then curious to search for some hidden topological features in understanding the eigenstates of $\hat{a}$.

To obtain the bulk topology of the system, we need to define the fictitious PBC. If we assume the dimension of Fock space to be $(D+1)$, a PBC can be implemented by linking the last site to the first site, i.e. $\hat{a}_{PBC,D}=\sum_{n=0}^{D}t_{j}\hat{c}_{j}^{\dagger}\hat{c}_{\text{mod}(j+1,D)}$. The characteristic equation of the annihilation operator, $\det(\hat{a}_{PBC,D}-EI)=0$, is expanded as $(-E)^{D+1}-\sqrt{(D+1)!}=0$. Employing the Stirling's formula, $(D+1)!\approx\sqrt{2\pi(D+1)}(\frac{D+1}{e})^{D+1}$, we derive $E_{l}/\sqrt{D}\approx\exp[-1/2+i(2l+1)\pi]$ with $\{l\in\mathbb{Z}|0\leq l\leq D\}$. Therefore, the spectrum of $\hat{a}$ under the PBC is a symmetric spectral loop with radius of $\sqrt{D/e}\rightarrow \infty$ surrounding the origin of the complex-plane, hence it encodes nontrivial point-gap topology~\cite{bergholtz2019exceptional}.
Similarly, we introduce the twisted boundary condition and define the spectral winding number as
\begin{equation}
W(\Omega)=\lim_{D\rightarrow \infty}\frac{1}{2i\pi}\int_{0}^{2\pi}d\theta\frac{\partial}{\partial\theta}\text{Tr}[\ln(\hat{a}_{D,PBC}(\theta)-\Omega)],
\tag{S3}
\end{equation}
where $\hat{a}_{D,PBC}(\theta)=e^{-i\theta/(D+1)}\sum_{j=0}^{D}t_{j}\hat{c}_{j}^{\dagger}\hat{c}_{\text{mod}(j+1,D)}$ is the form of $\hat{a}$ under the twisted boundary condition with twisted angle $\theta$, and $\Omega$ is the reference energy (point gap)~\cite{gong2018topological}. The winding is nontrivial because a varying $\theta$ smoothly rotate the spectral loop around the origin clockwise with a period of $2\pi$, yielding $W(\Omega)=-1$ for any $\Omega$ in the limit of $D\rightarrow\infty$.   Note also that in this limit,  the PBC spectral loop tends to cover the whole complex plane.

The presence of a nonvanishing spectral winding number $W(\Omega)$ implies that edge modes with eigenenergies $E=\Omega$ may emerge under appropriate boundary condition. We search for the edge modes under the natural sIBC of Fock space by solving the eigenequation $\hat{a}|\psi\rangle=\Omega|\psi\rangle$ with the condition $\lim_{j\rightarrow\infty}\psi_{j}=0$, where $\psi_{j}=\langle j|\psi\rangle$. Indeed it gives rise to localized solutions $|\psi\rangle=e^{-|\Omega|^2/2}\sum_{j=0}^{\infty}\frac{\Omega^{j}}{\sqrt{j!}}|j\rangle$.
Being the eigenstates of the annihilation operator, these edge modes are just, as expected,  the celebrated coherent states. Unlike ordinary edge modes that pile up at at the boundary, $|\psi\rangle$ is peaked at around $j\sim |\Omega|^2$ due to the inhomogeneity in the hopping strength $t_{j}$.  These observations are also similar to our results in the main text.

With this new interpretation of coherent states, we may further check the theory of BBC associated with point-gap topology. For example, the degeneracy of topological edge modes is typically linked to the absolute value of the above-defined topological invariant. For operator $\hat{a}^{n}$, since $\hat{a}^{p}|\psi\rangle=E^p|\psi\rangle$ once $\hat{a}|\psi\rangle=E|\psi\rangle$, the argument phase of PBC spectra of $\hat{a}^{p}$ is $p$ times of that of $\hat{a}$'s PBC spectra and then the winding number of $\hat{a}^{p}$ is $-p$. This then leads to the prediction that the topological edge modes of $\hat{a}^p$ is $p$-fold degenerate. It is indeed the case, since the set of coherent state $\{||\alpha|^{1/p}e^{i(\varphi+2q\pi)/p}\rangle, q=0,1,2\cdots (p-1)\}$ of $\hat{a}$ gives rise to the same eigenvalue of $\hat{a}^{p}$, with $\alpha=|\alpha|e^{i\varphi}$.
\end{widetext}

%\bibliographystyle{apsrev4-1}
%\bibliographystyle{abbrv}
%\bibliography{FST_ref}

\begin{thebibliography}{76}
\bibitem [{poi()}]{point_gap}%
  \BibitemOpen
  \href@noop {} {\ }\bibinfo {note} {The point gap in non-Hermitian systems
  means that the spectra do not touch a certain reference point.}\BibitemShut
  {Stop}%
\bibitem [{ima()}]{imaginary_gauge_field}%
  \BibitemOpen
  \href@noop {} {\ }\bibinfo {note} {Unlike the real gauge field, from which
  particles gain real phase $e^{\pm i\phi}$, the imaginary gauge field yields
  imaginary gauge phase $e^{\pm\phi}$ and hence asymmetric hopping. The
  energies of right- and left-going waves have opposite imaginary parts and the
  spectra take a loop shape on the complex plane and hence the point-gap
  topology.}\BibitemShut {Stop}%
\bibitem [{PBC()}]{PBCspectra}%
  \BibitemOpen
  \href@noop {} {\ }\bibinfo {note} {For example, the spectra of chain $\nu=0$
  are well fitted with an ellipse equation:
  $E_{\text{PBC},0l}^{(D)}/D\kappa=-\eta\sin\varphi_l+i(\xi\cos \varphi_l-1/2)$
  with $\xi=(2+D)/2D>\eta$ and $\varphi_l=2l\pi/(D+1)$.}\BibitemShut {Stop}%
\bibitem [{not()}]{note_integral}%
  \BibitemOpen
  \href@noop {} {\ }\bibinfo {note} {Since $|R_{\nu}^{(D)}(E)|=1$ on the
  spectral loops, the non-analytical point $|R_{\nu}^{(D)}(E)|=0$ isn't
  involved here.}\BibitemShut {Stop}%
\end{thebibliography}

%\begin{thebibliography}%

%

\end{document}